\documentclass[letterpaper,english,reprint, prb, superscriptaddress, nofootinbib, showpacs]{revtex4-1}
\usepackage[latin9]{inputenc}
\usepackage{textcomp}
\usepackage{amsmath}
\usepackage{amssymb}
\usepackage{graphicx}
\usepackage{esint}

\makeatletter

\DeclareRobustCommand{\greektext}{%
  \fontencoding{LGR}\selectfont\def\encodingdefault{LGR}}
\DeclareRobustCommand{\textgreek}[1]{\leavevmode{\greektext #1}}
\DeclareFontEncoding{LGR}{}{}
\DeclareTextSymbol{\~}{LGR}{126}

\makeatother

\usepackage{babel}
\begin{document}

\title{Influence of magnetic surface anisotropy on spin wave reflection
from the edge of ferromagnetic film}

\author{P.~Gruszecki}

\email{pawel.gruszecki@amu.edu.pl}

\affiliation{Faculty of Physics, Adam Mickiewicz University in Pozna\'{n}, Umultowska
85, Pozna\'{n}, 61-614, Poland }

\author{Yu.~S.~Dadoenkova}

\affiliation{Donetsk Physical and Technical Institute of the National Academy
of Sciences of Ukraine, 83114 Donetsk, Ukraine}

\affiliation{Ulyanovsk State University, 42 Leo Tolstoy str., 432000, Ulyanovsk,
Russian Federation}

\author{N.~N.~Dadoenkova}

\affiliation{Donetsk Physical and Technical Institute of the National Academy
of Sciences of Ukraine, 83114 Donetsk, Ukraine}

\affiliation{Ulyanovsk State University, 42 Leo Tolstoy str., 432000, Ulyanovsk,
Russian Federation}

\author{I.~L.~Lyubchanskii}

\affiliation{Donetsk Physical and Technical Institute of the National Academy
of Sciences of Ukraine, 83114 Donetsk, Ukraine}

\author{J.~Romero-Vivas}

\affiliation{Department of Electronic and Computer Engineering, University of
Limerick, Limerick, Ireland}

\author{K.~Y\@.~Guslienko}

\affiliation{1Depto. Física de Materiales, Facultad de Química, Universidad del
País Vasco, UPV/EHU, 20018 San Sebastián, Spain}

\affiliation{IKERBASQUE, The Basque Foundation for Science, 48013 Bilbao, Spain}

\author{M.~Krawczyk}

\email{krawczyk@amu.edu.pl}

\affiliation{Faculty of Physics, Adam Mickiewicz University in Pozna\'{n}, Umultowska
85, Pozna\'{n}, 61-614, Poland }

\date{\today}
\begin{abstract}
We study propagation of the Gaussian beam of spin waves and its reflection
from the edge of thin yttrium-iron-garnet film with in-plane magnetization
perpendicular to this edge. We have performed micromagnetic simulations
supported by analytical calculations to investigate the influence
of the surface magnetic anisotropy present at the film edge on the
reflection, especially in the context of the Goos-Hänchen effect.
We have shown the appearance of a negative lateral shift between reflected
and incident spin wave beams' spots. This shift is particularly sensitive
to the surface magnetic anisotropy value and is a result of the Goos-Hänchen
shift which is sensitive to the magnitude of the anisotropy and of
the bending of the spin wave beam. We have demonstrated that the demagnetizing
field provide graded increase of the refractive index for spin waves,
which is responsible for the bending. 
\end{abstract}

\maketitle
In recent years magnetic nanostructures with controlled magnetization
dynamics have been considered as candidates for design of new miniaturized
devices with enhanced performance and functionality for various applications,
e.g. heat transport, energy conversion, magnetic field sensing, information
storage and processing.\citep{Shibata04,Kajiwara13,Stamps14} Spin
waves (SWs), being propagating collective excitations of the magnetization
are also regarded as information carriers, which can be exploited
for information processing in devices potentially competitive with
standard CMOS systems.\citep{Krawczyk14,Chumak14} Thus, understanding
of SW properties in nanostructures is crucial in designing magnonic
units and this is one of the main goal in the research field called
magnonics.\citep{Kruglyak10,Demokritov13} It is expected that magnonic
devices allow energy-efficient processing of information which will
combine the advantages of photonics (high frequency and wide band)
and electronics (miniaturization) in a single unit.\citep{Khitun12}
One of the basic phenomena connected with wave propagation is the
wave transmission and reflection.\citep{Kostylev07,Chumak09,Dvornik11}
The reflection of SWs is determined by magnetic properties of the
film and boundary conditions at the border of the ferromagnetic material.
The reflection of SWs has already been investigated in theoretical
and experimental papers\citep{Chumak09,Kim08} where SWs were treated
as plane waves. Use of wave beams, instead of the plane waves or spherical
waves, in many cases, can be much more useful and opens new possibilities
due to its coherence and low divergence. The known example of the
wave beam is a light beam emitted by laser. Usually, its intensity
profiles can be described by Gaussian distribution (beams with such
property are called Gaussian beams). However, in magnonics the idea
of SW beams is unexplored, with only a few theoretical and experimental
studies considering formation of SW beams at low frequencies due to
the caustic or nonlinear effects.\citep{Vugal88,Floyd84,Schneider10,Gieniusz13,Kostylev11,Boyle96,Bauer97}

An interesting phenomenon characteristic for the reflection of beam
is a possibility for occurrence of a lateral shift of the beam spot
along the interface between the reflected and the incident beams -
this phenomenon is called as the Goos-Haenchen (GH) effect. The GH
effect was observed for electromagnetic waves,\citep{Goos47} acoustic
waves,\citep{Declercq08} electrons \citep{Sharma11} and neutron
waves.\citep{deHaan10} Also for SWs this topic was investigated theoretically
for the reflection of the exchange SWs (i.e., high frequency SWs with
neglected dipole-dipole interactions) from the interface between two
semi-infinite ferromagnetic films.\citep{Dadoenkova12} It was shown
that for the observation of the GH shift an interlayer exchange coupling
between materials is crucial. Recently, we analyzed the GH shift at
reflection of the SW's beam from the edge of the magnetic metallic
(Cobalt and Permalloy) and magnetic dielectric yttrium-iron-garnet
(YIG, $\mathrm{Y_{3}Fe_{5}O_{12}}$) films.\citep{Gruszecki14} We
showed that the GH effect exists for dipole-exchange spin waves and
can be observed experimentally. The magnetic properties at the film
edge were shown to be crucial for a shift of the SW's beam.

In this paper we analyze the SW beam reflected from the edge of the
thin ferromagnetic film. We focus our study on the magnetic properties
of the film's edge and its contribution to the shift of the SW beam.
We show, that measurements of this shift can provide information about
the local values of the surface magnetic anisotropy, and thus also
about the local magnetic properties at the edges of the magnetic film.
Our attention is concentrated on detailed investigation of the SW
reflection from the YIG film, a dielectric magnetic material highly
suitable for magnonic applications due to its low SW damping, which
is the smallest among all known magnetic materials.\citep{Serga12}
Recent experiments have shown possibility of fabrication of very thin
YIG films (with thicknesses down to tens of nm\citep{Sun12,Sun13}),
which can be patterned on nanoscale and in which the SW dynamics can
be controlled with metallic capping layers.\citep{Liu14,Pirro14}
The magnetic properties of the film edge influence SW dynamics, their
significance increases with decreasing size of device and will play
important role in spintronic and magnonic nanoscale devices.\citep{Shaw08,Putter09,Ozatay08,Xiao12,daSilva13}
However, edge properties at this scale are hardly accessible to experimental
techniques. In this paper, we propose a tool for the investigation
of the magnetic properties at the edges of thin ferromagnetic film,
which exploits shift of SW beams' spot at the reflection.

The micromagnetic simulations (MMS) and the analytical model of the
GH shift are described in section~\ref{sec:2Model-and-methods}.
Comparison of the results emerging from the analytical model with
MMS, development of the model of SW bending and the discussion of
the results are presented in section~\ref{sec:Results-and-discussion}.
The paper is summarized in section~\ref{sec:Conclusions}.

\section{Model and methods\label{sec:2Model-and-methods}}

\subsection{Model\label{sub:2_1Model}}

\begin{figure}[!ht]
\includegraphics[width=8cm]{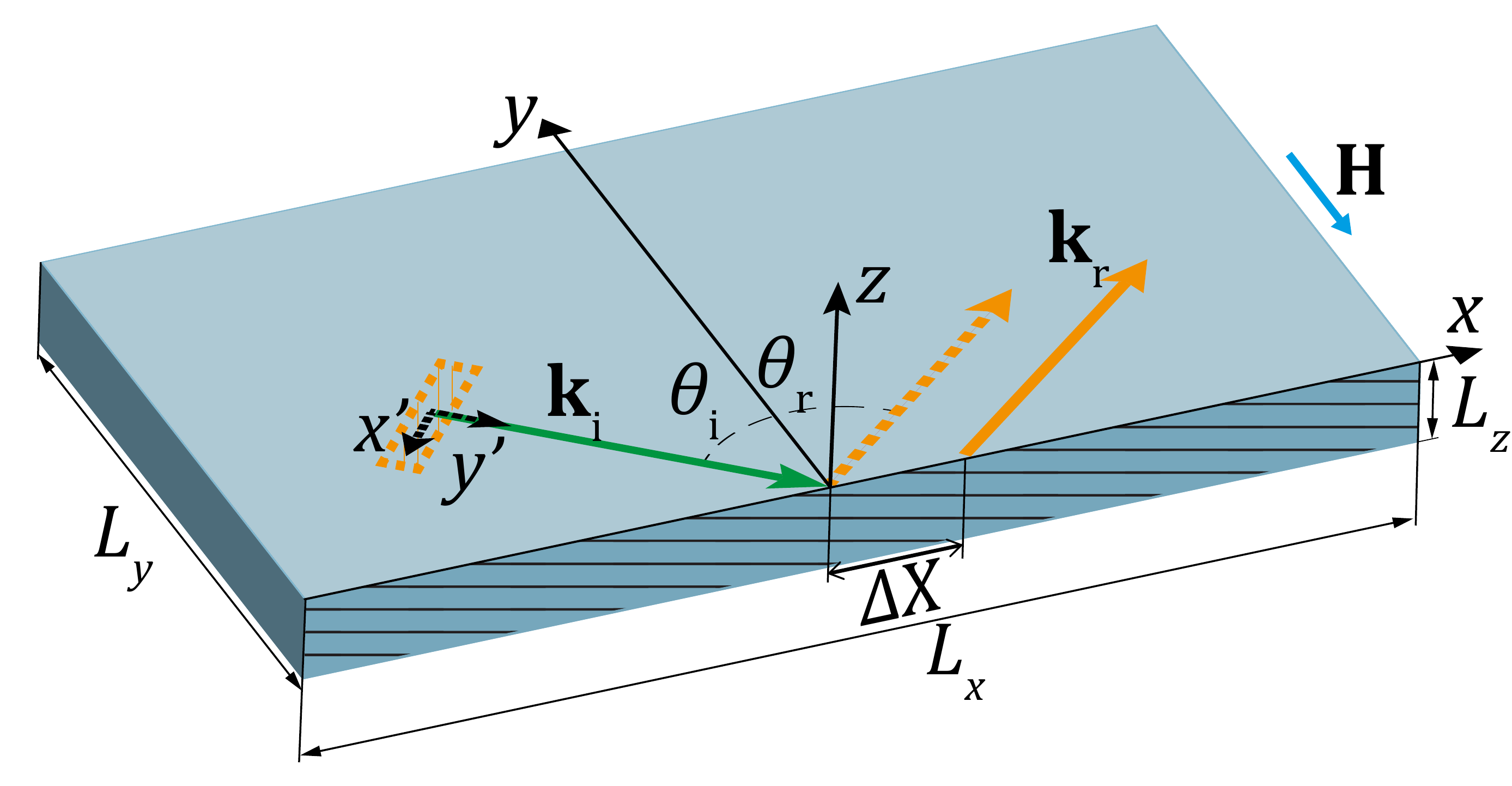} \protect\protect\caption{Schematic plot of the thin YIG film geometry considered in the paper.
The film has thickness $L_{z}$, which is much smaller than the film's
lateral sizes, $L_{x}$ and $L_{y}$. The ($x,y,z$) coordinating
system defines the structure with the film edge at $y=0$ (hatched
area). The coordinating system ($x',y',z$) defines the SW beam, with
the wave vector parallel to $y'$ and wave fronts parallel to $x'$.
The area hatched by orange lines and located in the center of coordinate
system ($x',y',z$) corresponds to the excitation area. $\mathbf{k_{\mathrm{i}}}$
and $\mathbf{k_{\mathrm{r}}}$ are wavevectors of incident and reflected
SW beams, respectively. $\Delta X$ is a total shift of the SW beam
reflected at the edge. \label{fig:F1} }
\end{figure}

We consider a thin YIG film with the thickness, $L_{z}=5$~nm, much
smaller than lateral dimensions of the film ($L_{z}\ll L_{x},L_{y}$)
as it is shown in figure~\ref{fig:F1}. The film is magnetically
saturated by an in-plane static external magnetic field $\mathbf{H}$
(we assume a value $\mu_{0}H=0.7$~T) which is applied along the
$y$-axis, perpendicular to the edge of the film. We study SWs which
propagate in the film plane ($x,y$). The considered edge of the film
is along $x$ axis and located at $y=0$. For description of the SW
propagation, it is more convenient to define also the second coordinate
system ($x',y'$). As it is shown in Fig.~\ref{fig:F1}, in this
coordinate system the wave vector $\mathbf{k_{\mathrm{i}}}$ of the
incident SW is parallel to $y'$ axis and wave fronts are parallel
to $x'$ axis. Therefore, we can define the angle of incidence $\theta_{\mathrm{i}}$
as the angle spanned between $\mathbf{k_{\mathrm{i}}}$ and normal
to the edge ($y$ axis). We limit angle of incidence to the value
$\theta_{\mathrm{i}}=60^{\circ}$ in this study. We assume the SW
frequency $f=35$~GHz, at this frequency the propagation is almost
isotropic in the film plane due to significant contribution of the
exchange interactions (as confirmed latter in the paper with calculated
isofrequency contours). In calculations we have used magnetic parameters
for YIG at low temperatures: saturation magnetization $M_{\mathrm{S}}=0.194\times10^{6}$~A/m
and exchange constant $A=0.4\times10^{-11}$ J/m. An additional advantage
of the YIG film is its relatively small static demagnetizing field,
which is proportional to $M_{\mathrm{S}}$. All these properties of
YIG simplify the analysis and helps us to focus mainly on the influence
of the surface magnetic anisotropy on the reflection of SWs. The surface
magnetic anisotropy can have different origin, besides change of the
crystallographic structure at the edge, the applying coating material
and roughness can also influence surface anisotropy.\citep{Yen79,Bruno89,Gradmann86,Johnson96,Jamet04}
Nonetheless, the microscopic mechanism of surface magnetic anisotropy
is not the subject of this paper, our main concern is the influence
of anisotropy on the reflection of a SW beam. 

Magnetization dynamics is described by the Landau-Lifshitz (LL) equation
of motion for the magnetization vector $\mathbf{M}$: 
\begin{equation}
\frac{\mathrm{d}\mathbf{M}}{\mathrm{d}t}=-\frac{\left|\gamma\right|}{1+\alpha^{2}}\mathbf{M}\times\mathbf{H}_{\mathrm{eff}}-\frac{\alpha\left|\gamma\right|}{M_{\mathrm{S}}\left(1+\alpha^{2}\right)}\mathbf{M}\times\left(\mathbf{M}\times\mathbf{H}_{\mathrm{eff}}\right),\label{eq:LLE}
\end{equation}
where: $\alpha$ - is the damping parameter, $\gamma$ - the gyromagnetic
ratio, $\mathbf{H}_{\mathrm{eff}}$ - effective magnetic field. The
first term in the LL equation describes precessional motion of the
magnetization around the effective magnetic field and the second term
enriches that precession by damping. The effective magnetic field
in general can consist of many terms. In this paper we consider only
the most important contributions: the external magnetic field $\mathbf{H}$,
the non-uniform exchange field $\mathbf{H}_{\mathrm{ex}}$ and the
long-range dipolar field $\mathbf{H}_{\mathrm{d}}$: $\mathbf{H}_{\mathrm{eff}}=\mathbf{H}+\mathbf{H}_{\mathrm{ex}}+\mathbf{H}_{\mathrm{d}}$.

\subsection{Analytical model of the GH shift\label{sub:2_2Analytical-model-of}}

In our analytical study we consider SWs with large wavevectors, where
the contribution from the dynamic dipole interactions is small and
SW dynamics is mainly determined by the exchange interactions. In
Eq. (\ref{eq:LLE}) we will also make a linear approximation, which
allows us to decompose the magnetization vector into a static part
equal to the saturation magnetization ${\mathbf{M}}_{\mathrm{S}}$
and a dynamical part laying in the plane perpendicular to the direction
of ${\mathbf{M}}_{\mathrm{S}}$: $\mathbf{M}=-M_{\mathrm{S}}\hat{\mathbf{y}}+\mathbf{m}\left(x,y,t\right)$.
This approximation is valid when the dynamical part of the magnetization,
$|\mathbf{m}|$ is much smaller than the saturation magnetization
$M_{\mathrm{S}}$. With this approximation we can assume harmonic
time dependency for $\mathbf{m}\propto e^{\mathrm{i}\omega t}$, where
$\omega$ is the angular frequency of SW. SWs' damping is neglected
here. To study the incidence and reflection of SWs from the edge of
the thin film, we start with dispersion relation for SWs in thin film
assuming that the wave vector is in plane of the film:\citep{Kalinikos86}
\begin{eqnarray}
\omega^{2} & = & \left[\omega_{\mathrm{H}}+l_{\mathrm{ex}}^{2}\omega_{\mathrm{M}}k^{2}+\omega_{\mathrm{M}}\left(1-f\left(kL\right)\right)\right]\times\nonumber \\
 &  & \left[\omega_{\mathrm{H}}+l_{\mathrm{ex}}^{2}\omega_{\mathrm{M}}k^{2}+\omega_{\mathrm{M}}f\left(kL\right)\sin^{2}\theta_{\mathbf{k}}\right],\label{eq:omega2}
\end{eqnarray}
where: $\mu_{0}$ is permeability of vacuum, $\omega_{\mathrm{H}}=\left|\gamma\right|\mu_{0}H_{\mathrm{i}}$,
is the internal field, for the in-plane magnetic field we assume $H_{\mathrm{i}}=H$ , $\omega_{\mathrm{M}}=\gamma\mu_{0}M_{\mathrm{S}}$,
and the exchange length $l_{\mathrm{ex}}=\sqrt{2A/(\mu_{0}M_{\mathrm{S}}^{2})}$,
$k$ is a wave number, $\theta_{\mathbf{k}}$ is the angle between
the saturation magnetization and the in-plane wave vector ${\bf {k}}$,
and 
\begin{equation}
f\left(x\right)=1-\frac{\left(1-\text{e}^{-x}\right)}{x}.\label{eq:fl}
\end{equation}
We assume that the wave number is large, i.e., the condition $kL_{z}\gg1$
is satisfied. Therefore, the SW dispersion relation can be simplified
($\theta_{\mathbf{k}}=\theta_{\mathrm{i}}$ in our geometry, see Fig.~\ref{fig:F1})
and reduced to the well-known Herring-Kittel equation: 
\begin{equation}
\omega^{2}=\left(\omega_{\mathrm{H}}+l_{\mathrm{ex}}^{2}\omega_{\mathrm{M}}k^{2}\right)\left(\omega_{\mathrm{H}}+l_{\mathrm{ex}}^{2}\omega_{\mathrm{M}}k^{2}+\omega_{\mathrm{M}}\sin^{2}\theta_{\mathrm{i}}\right),\label{eq:omega2}
\end{equation}
For further calculations we need formula for $k$ as a explicit function
of the frequency $\omega$. This dependence takes the following form:
\begin{equation}
k^{2}=\frac{\mu_{0}M_{\mathrm{S}}}{4A}\left(-2H-M_{\mathrm{S}}\sin^{2}\theta_{\mathrm{i}}+\sqrt{\frac{4\omega^{2}}{\mu_{0}^{2}\gamma^{2}}+M_{\mathrm{S}}^{2}\sin^{4}\theta_{\mathrm{i}}}\right).\label{eq:disp_k2}
\end{equation}

The calculation of the reflection coefficient requires the introduction
of the boundary conditions at the film edge (at $y=0$) for $\mathbf{m}$.
We show below that the magnetization vector $\mathbf{m}$ components
of the SW at the film edge (plane $y=0$) satisfy the following boundary
conditions of the Rado-Weertman type:\citep{Rado59} 
\begin{equation}
\left[\frac{\partial\mathbf{m}\left(y\right)}{\partial y}+d\mathbf{m}\left(y\right)\right]_{y=0}=0\label{eq:BCond}
\end{equation}
where $d$ is an effective pinning parameter. This pinning parameter
can take into account the dipole contribution in the finite width
stripe ($L_{y}$ is finite) and also contributions from the exchange
interaction and a surface magnetic anisotropy at the edge of the film.
The former contribution is anisotropic in restricted geometry and
have to be calculated in our particular case. To do that we start
from the boundary conditions written in the paper Ref.~{[}\onlinecite{Guslienko05}{]}
(see Eq.~(2) there):

\begin{equation}
\mathbf{M}\times\left(l_{\text{ex}}^{2}\frac{\partial\mathbf{M}}{\partial\mathrm{n}}-\frac{1}{\mu_{0}}\nabla_{\mathbf{M}}E_{\text{a}}+{\mathbf{H}}_{\text{d}}L_{z}\right)=0
\end{equation}
is the energy density of the uniaxial surface anisotropy, $\mathbf{n}$
is the unit vector along the anisotropy axis, $K_{\mathrm{s}}$ is
a surface anisotropy constant, and $\mathbf{H}_{\text{d}}$ is the
inhomogeneous magnetostatic field near the surface. The authors of
the papers {[}\onlinecite{Gus02,Guslienko05}{]} assumed that the
static magnetization is parallel to the magnetic element surface and
deduced the pinning of the dynamic magnetization component directed
perpendicularly to the surface. Such perpendicular magnetization component
corresponds to increasing of the surface magnetostatic energy, which
is an effective easy plane surface anisotropy. An additional uniaxial
surface anisotropy can be accounted leading to a re-normalization
of the magnetostatic anisotropy. The system tries to reduce the surface
magnetic charges developing some inhomogeneous magnetization configuration
near the surface. But this costs an additional volume magnetostatic
and exchange energy. The pinning was strong for thin ($L_{z}$ = 2--20
nm) magnetic elements reflecting a balance between these energy contributions.
The pinning calculated in Refs.~{[}\onlinecite{Gus02,Guslienko05}{]}
is a result of rapid change of the dynamic magnetostatic field near
the surface (on distance about of $L_{z}$). Whereas, in our case
the static magnetization is perpendicular to the surface (film edge,
the plane $y=0$) and the dynamical magnetization components are parallel
to the surface plane $xOz$. Moreover, there is a SW wave vector component
$k_{x}$ parallel to the surface. The pinning also has magnetostatic
contribution due to the strong dependence of the static dipolar field
$H_{\text{d}}\left(y\right)$ on the $y$-coordinate near the surface
plane $y=0$. This dipolar pinning competes with the pinning induced
by the surface anisotropy. Therefore, the contribution of even weak
surface anisotropy is important in this case. We can re-write the
boundary condition equations in the symmetric explicit form: 
\begin{equation}
\frac{l_{\text{ex}}^{2}}{L_{z}}\frac{\partial{\mathbf{m}}}{\partial{n}}+{\mathbf{h}}+\left[\frac{1}{2}-\frac{\kappa_{\text{s}}}{L_{z}}\right]{\mathbf{m}}=0,\label{eq:bc2}
\end{equation}
where $\kappa_{\text{s}}=2K_{\text{s}}/\mu_{0}M_{\text{S}}^{2}$,
$\mathbf{H}_{\text{d}}=\mathbf{H}_{\text{d}}^{o}+\mathbf{h}$ and
the contributions from exchange, dynamical dipolar, static dipolar,
and surface anisotropy fields, respectively, are included. The magnetization
precession in such non-ellipsoidal element under consideration (see
Fig.~\ref{fig:F1}) is elliptical due to non-equivalent dynamical
dipolar field components $h_{x}$ and $h_{z}$ along the film normal
$Oz$ and in-plane $Ox$ directions.

The dipolar field components can be expressed via the dynamical magnetization
components using the method of the tensorial magnetostatic Green functions,
see Appendix~\ref{sec:Green} and Ref.~{[}\onlinecite{Gus11}{]}.
Even assuming that the dynamical magnetization does not depend on
the thickness coordinate $z$ and averaging over $z$ we still have
two-dimensional problem because the magnetization depends on the in-plane
coordinates $x,y$. We assume that the dynamical magnetization can
be represented in the form $m_{\alpha}\left(x,y\right)=\text{exp}\left(ik_{x}x\right)m_{\alpha}\left(y\right)$,
where $k_{x}$ is almost continuous variable due to large element
size along $Ox$ direction. The magnetization profile $m_{\alpha}\left(y\right)$
can deviate from the plane wave near the element edge, where dipolar
fields are strongly inhomogeneous. Using the Green functions formalism
we simplify the boundary conditions given by Eq.~(\ref{eq:bc2})
and write them in the generalized Rado-Weertman form:

\begin{equation}
\left[\frac{\partial m_{\alpha}}{\partial y}+d_{\alpha}m_{\alpha}\right]_{y=0}=0,\;\text{where}\;\;\alpha=x,y
\end{equation}
but the pinning parameters $d_{x}$ and $d_{y}$ are different and
depend on the wave vector component $k_{x}$. We get 
\begin{eqnarray}
d_{x} & = & \frac{L_{z}}{l_{\text{ex}}^{2}}\left[\frac{1-f\left(k_{x}L_{z}\right)}{2}-\frac{\kappa_{\text{s}}}{L_{z}}\right],\nonumber \\
d_{z} & = & \frac{L_{z}}{l_{\text{ex}}^{2}}\left(\frac{f\left(k_{x}L_{z}\right)}{2}-\frac{\kappa_{\text{s}}}{L_{z}}\right).\label{eq:bc_d1}
\end{eqnarray}

The typical value of $k_{x}L_{z}$ is order of $1$ that corresponds
to wave number $k_{x}$ of about $0.1$ nm$^{-1}$, i.e. to the dipolar-exchange
SW regime. To simplify the further analytical consideration we assume
that the exchange energy dominates and neglect the dynamical dipolar
fields in the boundary conditions. Therefore the symmetric pinning
parameters $d=d_{x}=d_{z}$ are: 
\begin{equation}
d=\frac{L_{z}}{l_{\text{ex}}^{2}}\left(\frac{1}{2}-\frac{\kappa_{\text{s}}}{L_{z}}\right).\label{eq:bc_d2}
\end{equation}


Due to translational symmetry along the interface ($Ox$ direction),
the wave vector component along the interface ($k_{x}$) should be
conserved. As a consequence, the angle of incidence is equal to the
angle of reflection, $\theta_{\mathrm{i}}=\theta_{\mathrm{r}}$. Therefore,
based on Eqs.~(\ref{eq:disp_k2}) and (\ref{eq:BCond}) and assuming
$m_{\alpha}\left(y\right)$ dependence in the plane wave form we can
derive the equation for the reflection coefficient $R$ (see Appendix~\ref{sec:refl}):
\begin{equation}
R=\frac{\mathrm{i}\sqrt{k^{2}-k_{x}^{2}}+d}{\mathrm{i}\sqrt{k^{2}-k_{x}^{2}}-d}.\label{eq:ReflCoefR}
\end{equation}

When the incident SW is represented as a wave packet of a Gaussian
shape, with a characteristic length in momentum space $\Delta k_{x}'\ll k_{x}'$,
then according with the stationary phase method\citep{Dadoenkova12,Artmann48}
the reflected beam reveals a space shift relatively to the incident
wave packet of the length (the GH shift): 
\begin{equation}
\Delta X_{\mathrm{GH}}=-\frac{\partial\psi}{\partial k_{x}^{'}},
\end{equation}
where $\psi=\arctan\left(\Im(R)/\Re(R)\right)$ is the phase difference
between the reflected and incident waves, $\Im(R)$ and $\Re(R)$
are the imaginary and real parts of the reflection coefficient calculated
from Eq.~(\ref{eq:ReflCoefR}). Thus, the GH shift $\Delta X_{\mathrm{GH}}$
dependence on the magnetization pinning coefficient can be expressed
by the following equation: 
\begin{equation}
\Delta X_{\mathrm{GH}}=-\frac{2d\tan\theta_{\mathrm{i}}}{d^{2}+\left(k\cos\theta_{\mathrm{i}}\right)^{2}}.\label{eq:GH_d}
\end{equation}

\subsection{Micromagnetic simulations\label{sub:2_3Micromagnetic-simulations}}

Micromagnetic simulations (MMS) have been proved to be an efficient
tool for the calculation of SW dynamics in various geometries.
\citep{Venkat13,Kim11,Hertel04,Lebecki08} We have exploited an interface
with the GPU-accelerated MMS program MuMax3\citep{Vansteenkiste14}
which uses finite difference method to solve time-dependent LL Eq.~(\ref{eq:LLE}).

In our MMS we consider SWs propagation in thin-films and reflection
from the film edge. Simulations were performed for the system shown
in Fig.~\ref{fig:F1} of size $4000\times12000\times5$~nm ($L_{x}\times L_{y}\times L_{z}$),
which was discretized with cuboid elements of dimensions $2.5\times2.5\times5$~nm
($l_{x}\times l_{y}\times l_{z}$ are much less than $13$~nm, i.e.,
the exchange length of YIG). The surface magnetic anisotropy was introduced
in MMS by uniaxial magnetic anisotropy value $K_{\mathrm{u}}$ in
the single row of discretized cuboids at the film edge according with
the definition $K_{\mathrm{u}}=K_{\mathrm{s}}/l_{\mathrm{y}}$,\citep{footnote_1}
where $l_{\mathrm{y}}=2.5$~nm is a size of cuboid along normal to the edge.

\begin{figure}[!ht]
\includegraphics[width=8cm]{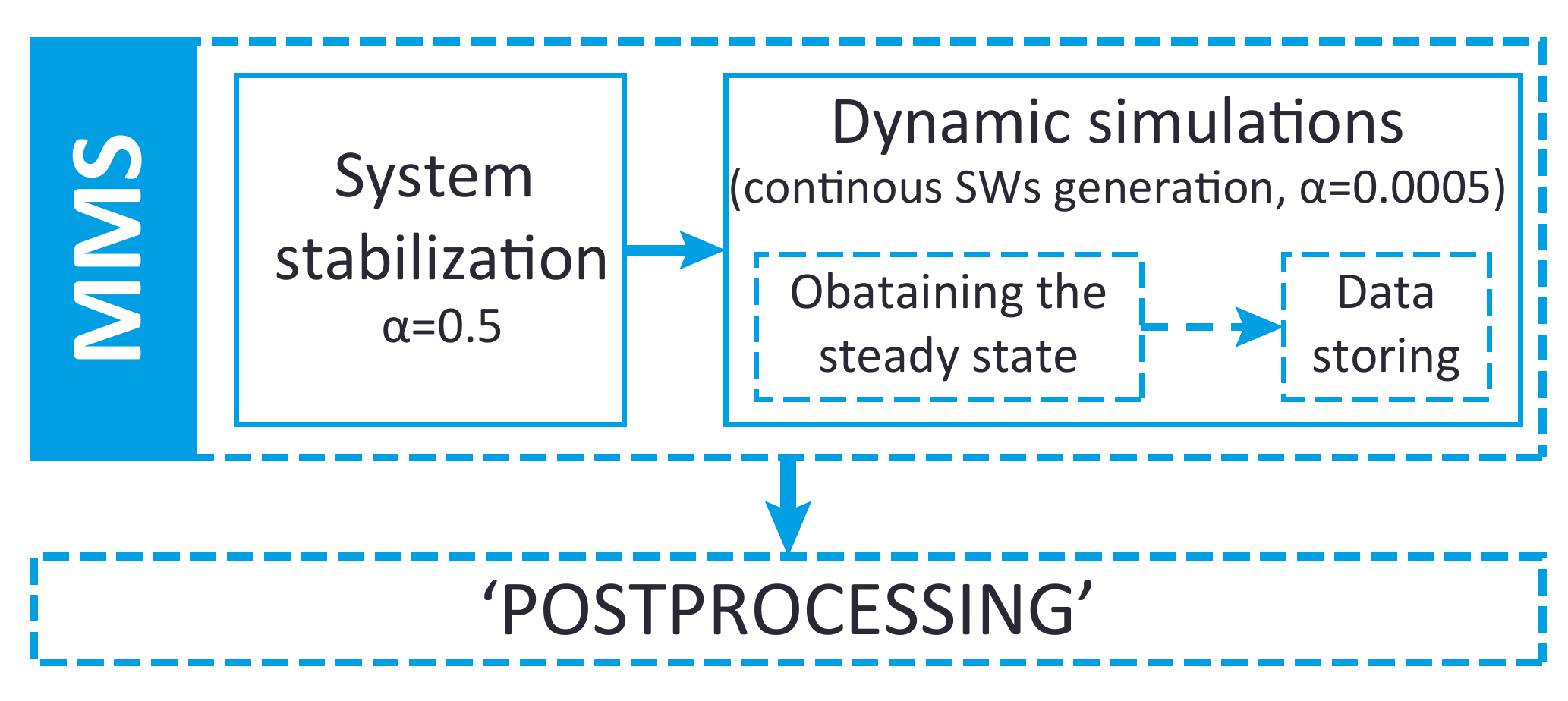} \protect\protect\caption{Algorithm of the SW dynamics calculations by MMS. MMS consist
of two steps. in the first step the system is stabilizing--the equilibrium
magnetic configuration is obtained. In the second step using the stabilized
magnetic configuration SWs are generated by applying small rf magnetic
field. The data stored during MMS are processed during the stage called
`POSTPROCESSING'--the final results of the SW dynamics are extracted.\label{fig:F2} }
\end{figure}

Simulations consist of two parts according to the algorithm presented
in Fig.~\ref{fig:F2}. First, we obtain the equilibrium static magnetic
configuration of simulated system. In this part of simulations we
start from random magnetic configuration in presence of high damping
($\alpha=0.5$). Then, the results of the first stage are used in
the dynamic part of simulations during which a SW beam is continuously
generated and propagates through the film with reduced, finite value
of damping parameter, $\alpha=0.0005$, comparable with the values
present in high quality YIG films.\citep{Serga12} The SWs are
excited in the form of a Gaussian beam. After sufficiently long time,
when incident and reflected beam are clearly visible and not changing
qualitatively in time the data necessary for further analysis (`POSTPROCESSING')
are stored. 

\begin{figure}[!ht]
\includegraphics[width=8cm]{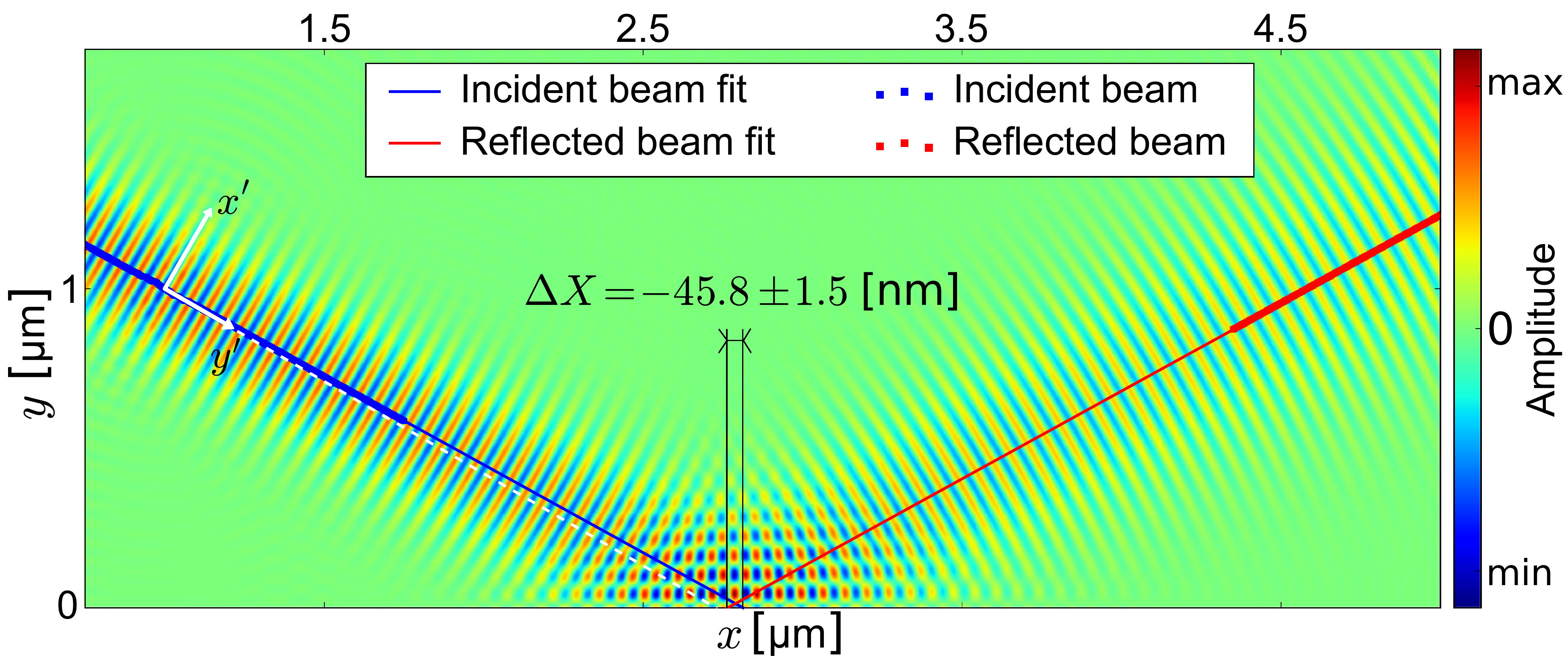} \protect\protect\caption{Exemplary result of MMS showing the color map of the dynamic magnetization
amplitude of the SW beam at reflection from the edge of the YIG film.
The presented result\textbf{ }was achieved for $K_{\mathrm{S}}=-0.1$~mJ/m\protect\protect\textsuperscript{2}.
At this value of the surface anisotropy the negative value of SW beam
shift is observed: $\Delta X=-45.8$~nm. The blue line corresponds
to the rays of the incident beam and red line to the reflected beam.
The dashed white line close to the ray of the incident beam points
at the direction of the SW wavevector.\label{fig:F3} }
\end{figure}

To generate Gaussian beam of SWs we introduce a narrow rectangular
area (excitation area, marked in Fig. \ref{fig:F1} with orange dashed
lines) with a long side parallel to the expected wave fronts (along
$x'$ axis). Within the excitation area we introduce a radio-frequency
(rf) magnetic field oscillating at frequency of $35$~GHz, $h_{\mathrm{dyn}}\left(x',y',t\right)=h_{0}(x{'},y')\exp\left(\mathrm{i}\omega t\right))$.
The field $h_{\mathrm{dyn}}$ is perpendicular to the static magnetic
field and its amplitude changes along the $x'$-axis according with
the Gauss distribution $G(x)=\exp\left[2\left(x{'}-x_{0}{'}\right)^{2}/\left(l\sigma\right)^{2}\right]$
and along $y'$-axis takes uniform, non-zero values only for small
window of width $w=5$ nm centered around $y_{0}{'}$ . $l$ is the
length of the excitation area (in our simulations $l=1.5$~\textgreek{m}m)
and $\sigma^{2}$ ($\sigma=0.2$) is a parameter which can be treated
as variance of the Gauss distribution centered around $x_{0}{'}$.
Hence, $h_{0}\left(x',y'\right)=h\Theta_{\mathrm{H}}(y{'}-y_{0}{'}+w/2)\Theta_{\mathrm{H}}(-y{'}+y_{0}^{'}+w/2)G(x{'})$,
where $\Theta_{\mathrm{H}}$ is the Heaviside step function. We assume
that $h=0.02H$ being maximum amplitude of the rf magnetic field which
needs to be small to stay in linear regime. Example result of MMS
is shown in Fig.~\ref{fig:F3}.

Extracting the precise value of the shift of SW beam $\Delta X$ from
MMS results requires the following three-stage procedure. In the first
stage we extract a series of SW intensity profiles using an array
of screen detectors parallel to the $y$-axis for different locations
($x_{j}$) along the film edge but far enough from the reflection
point, i.e., out of the interference pattern of SW near the reflection
point. At every point $x_{j}$ the intensity was calculated using
equation: $I_{x_{j}}\left(y\right)=\int_{0}^{4T}\left|m_{z}\left(x_{j},y,t\right)\right|\mathrm{d}t$,
where $T=1/f$ and $m_{z}$ is the component of the magnetization
vector perpendicular to the film plane. In the next stage, using Gaussian
fitting, we have extracted positions of centers of the intensity profiles
for every $y_{j}(x_{j})$. Having series of the peak positions and
its locations along $y$ (red and blue full dots in Fig.~\ref{fig:F3}
for the incident and reflected beams, respectively) we can extract
rays of the incident and reflected SW beams (red and blue solid line
in Fig.~\ref{fig:F3}, respectively). Finally, the value of the shift
$\Delta X$ can be easily calculated with small errors up to several
nanometers.

\section{Results and discussion\label{sec:Results-and-discussion}}

\begin{figure}[!ht]
\includegraphics[width=8cm]{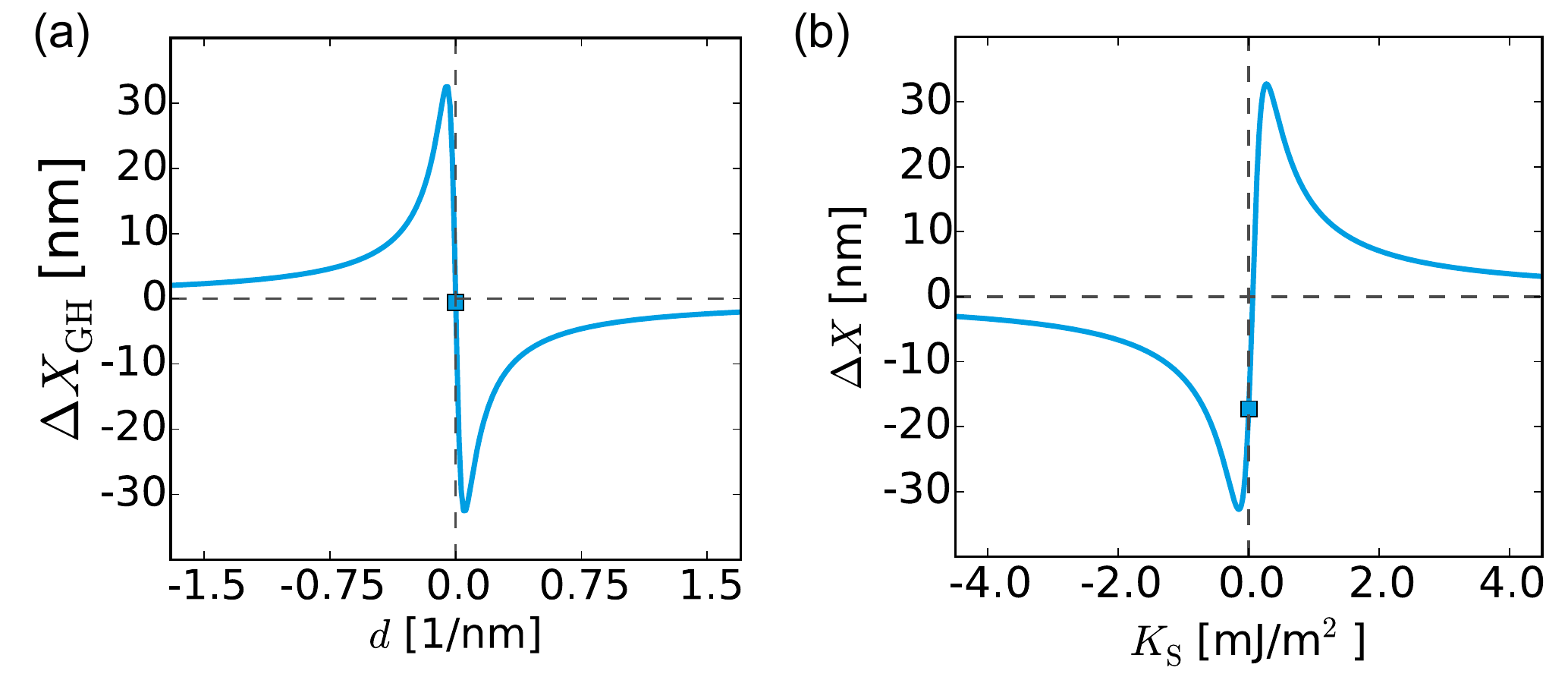} \protect\protect\caption{Analytical results of the GH shift in the reflection of the SW beam
from edge of the thin YIG film calculated using Eq. (\ref{eq:GH_d}) in dependence
on (a) pinning parameter $d$ and (b) magnetic surface anisotropy
constant $K_{\mathrm{s}}$. The blue squares corresponds to values
of GH shift for $d=0$ in (a) and $K_{\mathrm{s}}=0$ in (b).\label{fig:F4} }
\end{figure}

Result of the pure GH shift in dependence on the pinning parameter
obtained from the analytical model {[}Eq. (\ref{eq:GH_d}){]} is presented
in Fig.~\ref{fig:F4}(a). This dependence $\Delta X_{\mathrm{GH}}(d)$
is an antisymmetrical function with respect to $d=0$ and has maximum
and minimum value for $d<0$ and $d>0$, respectively. The effective
pinning parameter $d$ has contributions from the exchange interaction,
dipole interaction and magnetic surface anisotropy. To study influence
of the magnetic anisotropy we show also $\Delta X$ in dependence
on $K_{\mathrm{s}}$ in Fig.~\ref{fig:F4}(b) with solid line. GH
shift exists for $K_{\mathrm{s}}=0$ (marked with small square in Fig.~\ref{fig:F4}(b)) due to effective pinning coming
from the dipole interactions\citep{Guslienko05} and is $\text{\textgreek{D}}X_{\mathrm{GH}}=-17.0$~nm.
$\Delta X_{\mathrm{GH}}$ takes maximum absolute values for $K_{\mathrm{s}}=0.271$~mJ/m\textsuperscript{2}
and $K_{\mathrm{s}}=-0.153$~mJ/m\textsuperscript{2}. $\Delta X_{\mathrm{GH}}=0$
for $K_{\mathrm{s}}=0.59$~mJ/m\textsuperscript{2}, this is when
magnetic surface anisotropy compensate the effect of the dipole interactions
at the film edge. For large negative and positive values of $K_{\mathrm{S}}$
the GH shift tends monotonously to zero. This shows that the measure
of the GH shift can be used to indicate the surface magnetic anisotropy
at the thin film edge locally, especially in the range of its sudden
change, i.e., between $-0.153$ and $0.271$~mJ/m\textsuperscript{2}.
To test this possibility we perform MMS according with the procedure
described in section \ref{sub:2_3Micromagnetic-simulations}.

\begin{figure}[!ht]
\includegraphics[width=8cm]{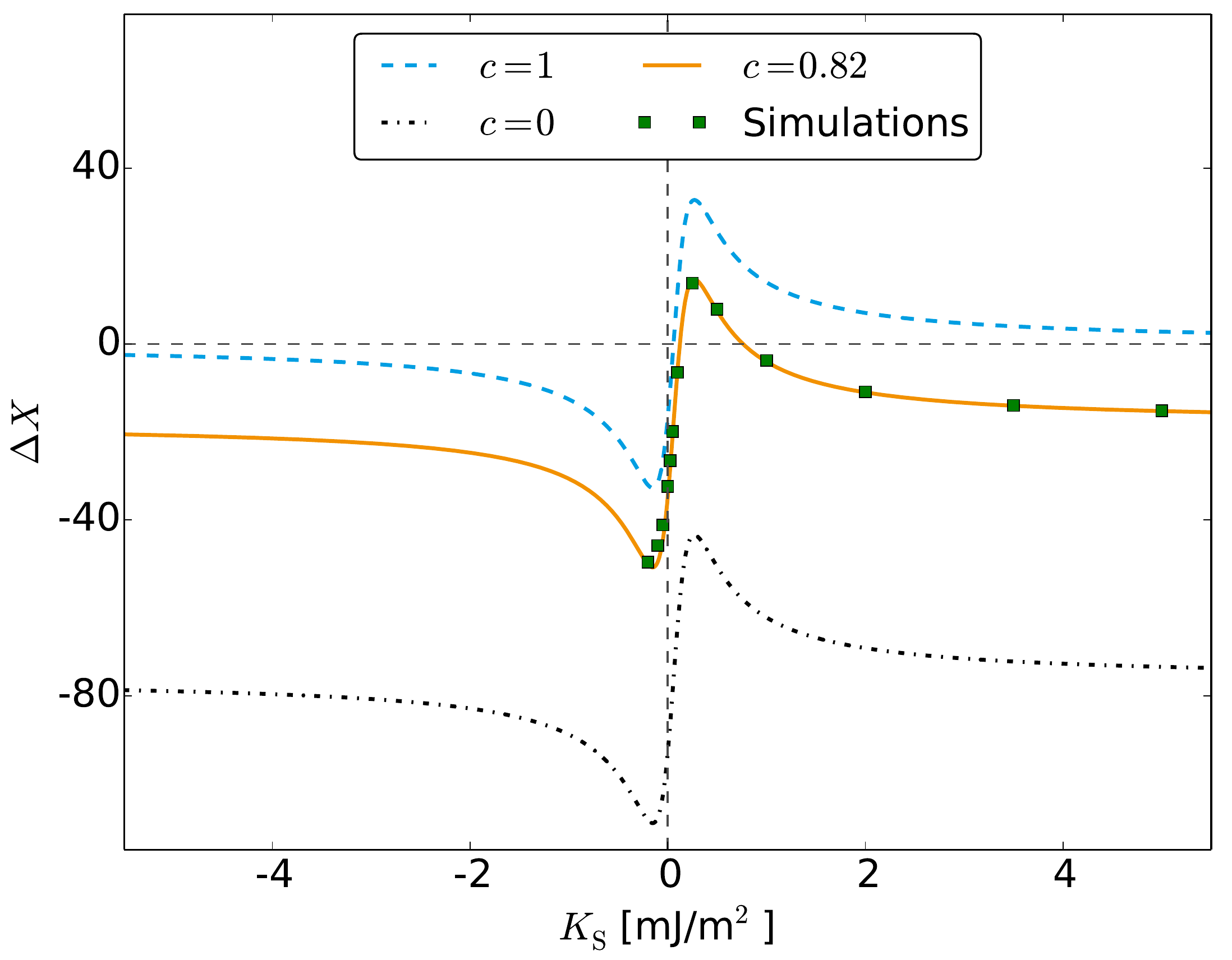} \protect\protect\caption{Results of the MMS (green solid points) and obtained from the analytical
models presenting dependence of the SW beam shift on the surface magnetic
anisotropy constant in the YIG film in the external field $0.7$~T.
Only the results for fully saturated sample $K_{\mathrm{s}}\ge K_{0}$
($K_{0}=-0.22$~mJ/m\protect\textsuperscript{2}) are shown. Dashed
blue line corresponds to basic analytical model for GH shift {[}Eq.
(\ref{eq:GH_d}){]} without included SWs bending corrections ($c=1$).
Dash-dotted black line presents results for the analytical model {[}Eq.~(\ref{Eq:k2_with_c}){]}
with $c=0$. Solid orange line correspond to results of the analytical
model with fitted $c=0.82$.\label{fig:F5} }
\end{figure}

Dependence of the SW beam shift on the surface anisotropy constant
obtained from MMS for $\mu_{0}H=0.7$~T in thin YIG film is presented
in Fig.~\ref{fig:F5} with green solid dots. The value of the shift
for $K_{\mathrm{s}}=0$ obtained from MMS is $\Delta X=-32.4$~nm
and this is significantly larger than the GH shift obtained from analytical
solutions ($\Delta X_{\mathrm{GH}}=-9$~nm). The maximal value of
$\Delta X=13.83$~nm is found for $K_{\mathrm{s}}=0.25$~mJ/m$^{2}$
and minimal is $\Delta X=-49.6$~nm for $K_{\mathrm{s}}=-0.2$~mJ/m$^{2}$,
i.e., out of the scale presented in Fig.~\ref{fig:F5}. This dependence
is quantitatively similar to the function obtained in the analytical
model (Fig.~\ref{fig:F4}). Nevertheless, there are distinct differences
between both results.

\begin{figure}[!ht]
\includegraphics[width=8cm]{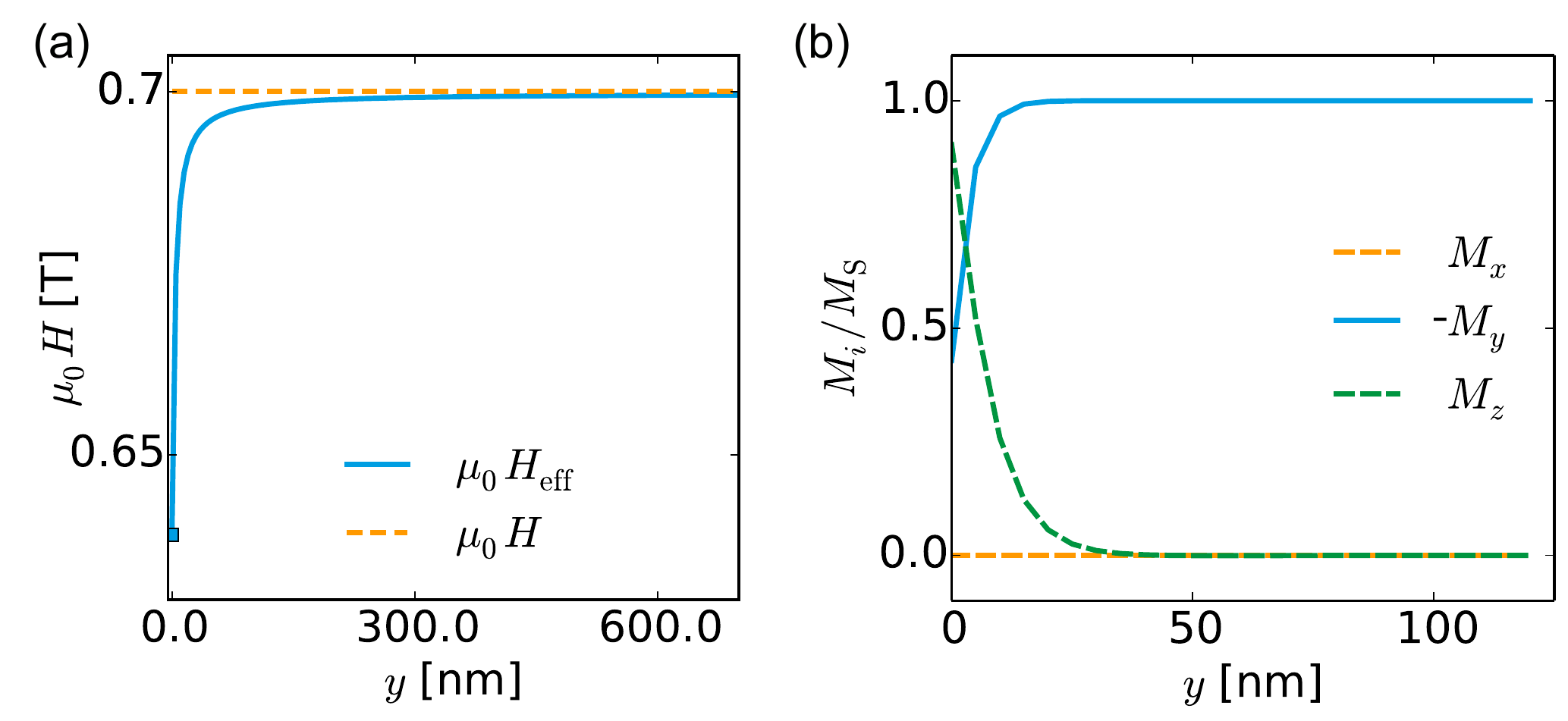} \protect\protect\caption{(a) Effective magnetic field along the $y$-axis inside YIG film for
$K_{\mathrm{s}}=0$ is shown with blue solid line, orange dashed line
corresponds to the external magnetic field (0.7 T). (b) Static magnetic
configuration in the vicinity of thin YIG film edge for $K_{\mathrm{s}}=-2$~mJ/m\protect\protect\protect\textsuperscript{2}
(i.e., $K_{\mathrm{s}}<K_{0}$); blue solid, orange dashed and green
dash-dotted line marks $x$, $y$ and $z$ component of the magnetization
vector normalized to unity, respectively.\label{fig:F6} }
\end{figure}

The analytical model is based on number of assumptions which are absent
in MMS, thus there is a disagreement between both results. The full
saturation of the magnetization is important assumption made in the
analytical model. Surface magnetic anisotropy at the edge of the film
increases (or decreases) a value of a total internal magnetic field
at this edge. This change of the internal magnetic field in the single
computational cell at the edge is equal to the surface anisotropy
field $\mu_{0}H_{\mathrm{s}}=2K_{\mathrm{s}}/\left(M_{\mathrm{s}}l_{y}\right)$.
Therefore, exactly at the film edge the demagnetizing field can be
compensated or enhanced by surface anisotropy field, in dependence
on the sign of $K_{\mathrm{s}}$. When surface anisotropy constant
$K_{\mathrm{s}}=K_{0}=-\frac{1}{2}\mu_{0}L_{z}M_{\mathrm{s}}\left(H-M_{\mathrm{s}}\right)\thickapprox-0.22$~mJ/m\textsuperscript{2}
the total internal field at the surface is equal $0$ (the demagnetizing
field at the edge is compensated by the anisotropy field). Hence,
for $K_{\mathrm{s}}<K_{0}$ the equilibrium orientation of the magnetization
at the edge of the film rotates from the saturation direction (this
is transformation from the easy axis to the easy plane configuration).
This is demonstrated in Fig.~\ref{fig:F6}\,(b), where magnetization
components along $x$, $y$ and $z$ axis are presented as a function of the
distance from the film edge for $K_{\mathrm{s}}=-2.0$~mJ/m\textsuperscript{2}.
We can see the rotation of the magnetization from the $y$ direction
(saturation direction) towards the $x$ axis. The change of the magnetization
configuration is not taken into account in the model developed in
Sec.~\ref{sub:2_2Analytical-model-of}. Moreover, in real sample
there is a possibility for appearing domain walls along the film edge,
which can create additional factor for complexity of the problem.
Therefore, in this paper we limit the analysis to the fully saturated
sample, i.e., when $K_{\mathrm{s}}\ge K_{0}$ (we note, that the exact
value of $K_{0}$ depends on the magnitude of the external magnetic
field).

The model shown in Sec.~\ref{sub:2_2Analytical-model-of} doesn't take into account influence of inhomogeneity of the internal magnetic field
on the behavior of propagating beam especially in the vicinity of
the film edge. It is probably the next most important factor, which
makes a comparison of the analytical and MMS results difficult for
the saturate state. In MMS, and in a real sample, the magnetization which is perpendicular
to the film edge creates an inhomogeneous static demagnetizing field,
which is directed opposite to the magnetization saturation. In result,
the internal magnetic field decreases monotonically when moving from
the film center towards its edge {[}solid-blue line in Fig.~\ref{fig:F6}(a){]}.
In our interpretation this inhomogeneity is responsible for a shift
of $\Delta X$ in MMS towards negative values as compared to the results
of Eq.~(\ref{eq:GH_d}) {[}see, Fig.~\ref{fig:F6}{]}: for very
high positive $K_{\mathrm{s}}$ the $\Delta X_{\mathrm{GH}}$ monotonously
tends to $0$ in the analytical model {[}Fig.~\ref{fig:F4}(b){]},
while results of the MMS show that the value of $\Delta X$ reaches
non-zero negative value even for very high value of $K_{\mathrm{s}}$
($\Delta X=-15.2$~nm for $K_{\mathrm{s}}=10$~mJ/m\textsuperscript{2}).
This means, that the inhomogeneity of the internal magnetic field
in the close vicinity of the thin film edge causes an increase of
the refractive index for SWs \citep{Kim08,Vansteenkiste14} and consequently
results in bending of the SW beam and changes the shift measured in
far field.

Here, we propose the simple analytical model of the wave bending,
which allows to estimate factor which correct the value of the GH
SW beam's shift obtained from Eq.~(\ref{eq:GH_d}).  We will consider gradual change
of the refractive index for SWs in the vicinity of the film edge.

\begin{figure}[!ht]
\includegraphics[width=8cm]{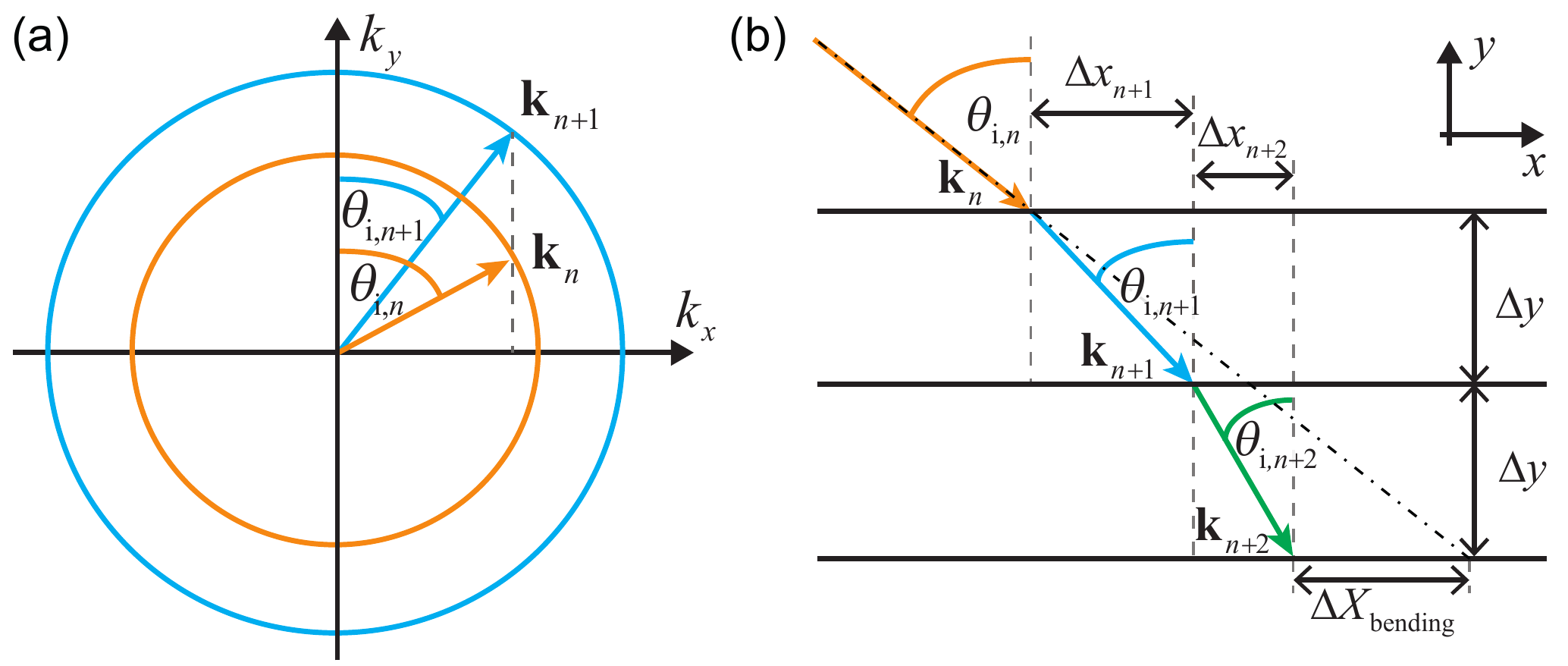} \protect\protect\caption{(a) Explanation of the refraction of the wave on the interface between
media with low refractive index ($n$-th slice, orange line) and high
refractive index ($n+1$ slice, blue line) based on isofrequency contours
analysis. The conservation of the $x$ components of the wavevectors
in the refraction is required by the translational symmetry along
the $x$. (b) Example of two refractions on the interfaces between
$n$-th and $n+1$, and between $n+1$ and $n+2$ slice. The refractive
index increases with increasing index $n$. The beam shift $\Delta X_{\mathrm{bending}}$
resulting from the bending is shown schematically.\label{fig:F7} }
\end{figure}

Similarly as in optics, the refraction law for SWs (i.e., Snell law)
can be concluded from analysis of the isofrequency contours and momentum
conservation of the wavevector component parallel to the edge ($k_{x}=\mathrm{const.}$).
Let us assume, that full distance of the gradual change of the refractive
index in thin film can be divided into $N$ thin slices, numbered
with integer $n$. Therefore, we can assume multiple refractions on
$(N-1)$ parallel planes, separating neighbor slices {[}Fig.~\ref{fig:F7}(b){]}.
At the plane between two arbitrary slices $n$ and $n+1$ according to
Snell low $\left(k_{n}\sin\theta_{\text{i},n}=k_{n+1}\sin\theta_{\text{i},n+1}\right)$
{[}Fig.~\ref{fig:F7}(a){]}. Thus, we can calculate final angle of
incidence after passing $N$ slices: $\sin\theta_{\mathrm{i},N}=\frac{k_{0}}{k_{N}}\sin\theta_{\mathrm{i},0}$,
where $k_{N}$ and $\theta_{\mathrm{i},N}$ are wavenumber and incident
angle in the last slice. $k_{0}\equiv k(y_{0})$ and $\theta_{\mathrm{i},0}$
are wavenumber and incident angle in the initial media, i.e., in the
interior far from the edge of the film. Rewriting the expression for
space coordinates (substitute $k(y)\equiv k_{N}$ and $\theta_{\mathrm{i}}(y)\equiv\theta_{\mathrm{i},N}$)
we obtain: 
\begin{equation}
\theta_{\mathrm{i}}(y)=\arcsin\left(\frac{k_{0}}{k(y)}\sin\theta_{\mathrm{i},0}\right).\label{eq:theta_i_y}
\end{equation}
From here, the final value of $\theta_{\mathrm{i}}(y)$ at the $y$
distance from the film edge can be calculated, if the value of $k(y)$,
initial values of the wavevector $k_{0}$ and the initial angle of
the incidence $\theta_{\mathrm{i},0}$ are known.

The knowledge of the incident angle at the vicinity of the film edge
$\theta_{\mathrm{i}}(y)$ allows us to derive formula which describes
propagation of the beam through the area of gradually changed refractive
index. Similarly to previous approach, let us consider single refraction
on the interface between $n$ and $n+1$ slice, as it is shown in
Fig.~\ref{fig:F7}(b). Projection of the beam path onto the direction
tangential to the interface is $\Delta x_{n+1}=\Delta y\tan\theta_{\mathrm{i},n+1}$
where $\Delta y$ is the thickness of the slice (we assume here for
simplicity slices of the same thickness). Generalizing this formula
for cascading refraction on $N$ slices we obtain: 
\begin{eqnarray}
x_{N} & = & \sum_{n=1}^{N}\Delta x_{n}=\Delta y\sum_{n=1}^{N}\tan\theta_{\mathrm{i},n}\nonumber \\
 & = & \Delta y\sum_{n=1}^{N}\tan\arcsin\left(\frac{k_{0}}{k_{n}}\sin\theta_{\text{i},0}\right)\label{Eq:xn_10}
\end{eqnarray}
Assuming infinitesimally small distance between interfaces $\Delta y\rightarrow dy$
we can transform summation into integration along the $y$ axis between
initial point of the SW generation $y_{0}$ and the point of the reflection
$y_{N}$: 
\begin{eqnarray}
x(y_{0},y_{N}) & = & \int_{y_{0}}^{y_{N}}\mathrm{d}y\tan\arcsin\left(\frac{k_{0}}{k_{n}}\sin\theta_{\text{i},0}\right)\nonumber \\
 & = & \int_{y_{0}}^{y_{N}}\mathrm{d}y\frac{\sin\theta_{\text{i},0}}{\sqrt{\left(\frac{k(y)}{k_{0}}\right)^{2}-\sin^{2}\theta_{\text{i},0}}}.\label{Eq:xn_11}
\end{eqnarray}
Therefore, if the edge where reflection takes place is located at
$y_{N}=0$ (like in Fig.~\ref{fig:F1}), the SW beam shift introduced
by bending and measured in far field (observation point is assumed
at the same distance from the edge as the source of the SW) is described
by following formula: 
\begin{equation}
\Delta X_{\mathrm{bending}}=2\left[y_{0}\tan\theta_{\mathrm{i}}-x(y_{0},0)\right],\label{eq:ShiftBending}
\end{equation}
where factor 2 is included to take into account the beam way from
source to the edge and after reflection to the observation point.

The total shift of the SW beam observed in far field is a sum of the
GH shift and the shift resulting from the bending: 
\begin{equation}
\Delta X=\Delta X_{\mathrm{bending}}+\Delta X_{\mathrm{GH}}.\label{eq:TotalShift}
\end{equation}
Eq.~(\ref{eq:TotalShift}) is a general formula describing a total
shift of the wave beam propagating in media with gradual change of
the refractive index. Way of the beam ray depends on relation describing
$\mathbf{k}(y)$. And this is main bottleneck in this approach: unknown
formula for wavevector of the SW in dependence on $y$ in an area
of the inhomogeneous effective magnetic field. However, knowledge
of the dispersion relation in homogeneous film, Eq.~(\ref{eq:disp_k2}),
can help in qualitative modeling of the total shift of the SW beam
also in a part of the film with gradual change of the refractive index.

Demagnetization field decreases a value of the internal magnetic field,
its dependence on distance from the edge can be expressed by the relation:
$H_{\mathrm{d}}(y)=4M_{\mathrm{S}}\arctan\left[L_{z}/(2y)\right]$.\citep{Joseph65}
This field far from the edge of the film tends to zero {[}Fig.~\ref{fig:F5}~(a){]}.
We can substitute bias magnetic field in Eq.~(\ref{eq:disp_k2})
with the internal field $H\rightarrow H(y)\equiv H-H_{\mathrm{d}}(y)$.
This approach should be valid for slow change of $H_{\mathrm{d}}(y)$
with distance. However, in the vicinity of the thin film edge we observe
a rapid change of the demagnetizing field {[}Fig. 6(a){]} and the
plane-wave approximation fails. Therefore, we propose to introduce
homotopic transformation of the demagnetizing field: $H(y)=H-\left[cH_{\mathrm{d}}(y_{0})+\left(1-c\right)H_{\mathrm{d}}(y)\right]$
with parameter $c\in\left[0,1\right]$, which will reduce influence
of the rapid changes of the demagnetizing field on wave vector. Then,
the wave vector magnitude can be described by the following equation:
\begin{eqnarray}
k^{2}(y,\theta_{\mathrm{i}}) & = & \frac{\mu_{0}M_{\mathrm{S}}}{4A}\Biggl(-2\left(H-\left[cH_{\mathrm{d}}(y_{0})+\left(1-c\right)H_{\mathrm{d}}(y)\right]\right)\nonumber \\
 &  & -M_{\mathrm{S}}\sin^{2}\theta_{\mathrm{i}}+\sqrt{\frac{4\omega^{2}}{\mu_{0}^{2}\gamma^{2}}+M_{\mathrm{S}}^{2}\sin^{4}\theta_{\mathrm{i}}}\Biggr),\label{Eq:k2_with_c}
\end{eqnarray}
where $y_{0}$ can be interpreted as position of the source of the
SW beam, i.e., the position where the demagnetizing field magnitude
is close to zero.

Now, the only issue is a correct choice of the parameter $c$ in Eq.~(\ref{Eq:k2_with_c})
to find $k(y)$, and to fit $\Delta X(K_{\mathrm{s}})$ dependence
to the curve obtained from MMS. Taking $c=1$ we assume, that isofrequency
contours are the same in whole sample ($k$ does not depend on $y$),
the wave propagate in homogeneous internal magnetic field equal $H$
{[}because $H_{d}(y_{0})\cong0${]} and Eq.~(\ref{Eq:k2_with_c})
reduces to Eq.~(\ref{eq:disp_k2}). In this case $\Delta X_{\mathrm{bending}}=0$
and $\Delta X=\Delta X_{\mathrm{GH}}$, i.e, the total shift is equal
to Eq.~(\ref{eq:GH_d}). The obtained curve is plotted in Fig.~\ref{fig:F5}
with blue dashed line. In opposite limit, $c=0$ the dependence $k(y)$
follows exactly the change of $H_{\mathrm{d}}(y)$. In this case the
$\Delta X_{\mathrm{total}}(K_{\mathrm{s}})$ calculated from Eq.~(\ref{eq:TotalShift})
is shown in Fig.~\ref{fig:F5} with black dotted line. It takes value
$\Delta X=\Delta X_{\mathrm{bending}}=-79$~nm for large $K_{\mathrm{s}}$,
much below the value obtained from MMS. This discrepancy exists, because
for each $y$ we took in calculation of the $\Delta X_{\mathrm{bending}}$
{[}in the integral Eq.~(\ref{Eq:xn_11}){]} the dispersion relation
Eq.~(\ref{Eq:k2_with_c}) which is for the film with homogenous magnetic
field. In real situation, the dispersion relation in the area of inhomogeneous
refractive index will be different from the local value. Thus averaging
of the demagnetizing field across some distance shall improve the
estimation. Moreover, the $c$ shall depend on the relative value
of the wavelength to the special changes of the refractive index.
Thus, the value of $c$ from Eq.~(\ref{Eq:k2_with_c}) needs to be
treated as a parameter, which includes an effective influence of the
inhomogeneity of the refractive index on the dispersion relation of
SWs. For $c=0.82$ we have obtained very good agreement between results
of MMS and analytical model $\Delta X(K_{\mathrm{s}})\approx\Delta X(K_{s})$ for $K_{\mathrm{s}}>-0.22$~mJ/m\textsuperscript{2}, as it is shown
in Fig.~\ref{fig:F6} with orange solid line.

The value of \textgreek{D}X is very sensitive for small change of
$K_{\mathrm{s}}$ between the extremes. The magnetic surface anisotropy
in ferromagnetic films can take different values, however the most
interesting is the range around $0$, where the transition from easy
axis into easy-plane anisotropy takes place. Thus, the measure of
the SW beam's shift can indicate the local surface magnetic anisotropy
at the film edge with spatial resolution limited by the size of the
width of the SW beam. This information shall be important also for
understanding and exploiting SW excitations and actuation at the surface
of YIG film being in contact with Pt, where the magnetic surface anisotropy
was shown to play a significant role in the spin pumping.\citep{Xiao12,daSilva13,Reshetnyak04,Zhou13}
Further investigation is required to test an influence of the second
ferromagnetic material attached to the YIG film edge on the SW reflection.
Here, extension of the analytical model with properly defined boundary
conditions will be required.\citep{Skarsvag14,Kruglyak14}

The micromagnetic simulations were conducted for the value of the damping parameter $\alpha=0.0005$, which is close to the value of thick\cite{Serga12} and also very thin films (tens of nm thick) of YIG as demonstrated experimentally recently.\cite{Jungfleisch15} We have performed additional simulations for slightly smaller and larger damping to check the influence of damping on GH shift. Apart from increase of the amplitude of the reflected SW beam with decreasing $\alpha$ the results of the GH shift are very close to those presented in the manuscript. This indicates that small changes of the homogeneous damping does not affect GH shift. However, for the electromagnetic waves reflected from the interface GH shift is strongly affected if the reflected material is a medium with strong absorption,\citep{Wild82,Chern14} where the absorption can even change the sign of the GH shift. Thus, we can suppose, that also in magnonics the inhomogeneous damping, especially with its high value at the interface or in the media behind the reflection edge, will influence GH shift. Further investigations are necessary to elucidate the role of inhomogeneous damping on the GH shift in the reflection of SWs from the edge or interface ferromagnetic films.

\section{Conclusions\label{sec:Conclusions}}
We have performed analytical and numerical study of the SWs beams
shift at the reflection from the edge of the YIG thin film in dependence
on the surface magnetic anisotropy present at the film edge. The GH
effect and SW bending are shown to contribute to the SW beam's shift
measured in far field. We have shown that the GH shift is modulated
in a broad range by changes of the surface magnetic anisotropy constant
between two extremes: $K_{\mathrm{s,min}}=-0.155$~mJ/m\textsuperscript{2}
and $K_{\mathrm{s,max}}=0.275$~mJ/m\textsuperscript{2}. It means
that even small change of the surface magnetic anisotropy, some imperfections
or changes in a surrounding of the film edge can result in significant
change in the reflected SW beams position in far field. The demagnetizing
field gradually changes the refractive index of SWs of the film with
its sudden increase near the film's edge. This variation of the refractive
index bends the SW propagating towards (and outwards) of the edge
and introduce additional shift of the SW beam. However, this shift
is independent on the surface magnetic anisotropy. For large positive
values of $K_{\mathrm{s}}$ the value of $\Delta X$ is almost insensitive
to changes of the magnitude of the magnetic surface anisotropy and
is mainly a result of the SW bending due to gradually increased refractive
index at the film edge. These results shall be of importance for magnonics,
its applications for sensing and also for developing a new direction
of research devoted to metamaterial properties for SWs, especially
the graded index magnonics.

\appendix

\section{Green functions}

\label{sec:Green} The $x$ and $z$ components of the magnetostatic
fields can be expressed as: 
\begin{eqnarray}
h_{x}(x,y)=\int dx'dy'G_{xx}(x,y;x',y')m_{x}(x',y'),\nonumber \\
h_{z}(x,y)=\int dx'dy'G_{zz}(x,y;x',y')m_{z}(x',y'),\label{eq:A1}
\end{eqnarray}
where 
\begin{eqnarray}
G_{\alpha\beta}(\mathbf{\rho};\mathbf{\rho}') & = & \frac{1}{(2\pi)^{2}}\int d^{2}\mathbf{k}G_{\alpha\beta}(\mathbf{k})\text{e}^{i\mathbf{k}\cdot(\mathbf{\rho}-\mathbf{\rho}')},\\
G_{xx}(\mathbf{k}) & = & -f(kL_{z})\frac{k_{x}^{2}}{k^{2}},\;\;G_{zz}(\mathbf{k})=-\left[1-f(kL_{z})\right],\nonumber 
\end{eqnarray}
and function $f$ is defined in Eq.~(\ref{eq:fl}).

Substituting $m_{\alpha}(x,y)=\exp{(ik_{x}x)}m_{\alpha}(y)$ to Eq.~(\ref{eq:A1})
and performing integration over $x'$ and $k_{x}$we can get the dipolar
field in the form 
\begin{equation}
h_{\alpha}(x,y)=\text{e}^{ik_{x}x}\int dy'g_{\alpha\alpha}(y,y')m_{\alpha}(y'),
\end{equation}
where the magnetostatic kernels are calculated using Ref.~{[}\onlinecite{Gradshteyn}{]}:
\begin{eqnarray}
g_{zz}(\eta)=\frac{1}{\pi L_{z}}\left[K_{0}\left(k_{x}\sqrt{L_{z}^{2}+\eta^{a}}\right)-K_{0}(k_{x}|\eta|)\right]\label{eq:gzz}
\end{eqnarray}
and 
\begin{eqnarray}
g_{xx}(\eta)=-\frac{1}{2}k_{x}\text{e}^{-k_{x}|\eta|},\;\;\text{at}\;\;k_{x}L_{z}\gg1,
\end{eqnarray}
above, $K_{0}$ is the modified Bessel's function of the zero order,
and $\eta=y-y'$. The $zz$-component of the magnetostatic kernel
in Eq.~(\ref{eq:gzz}) has logarithmic singularity at $\eta=0$.
The component $g_{xx}(\eta)$ has sharp maximum at $\eta=0$ because
of the condition $k_{x}L_{z}\gg1$. This justify using the Taylor
series decomposition near the point $\eta=0$ calculating the dynamical
dipolar fields and derivation of the boundary conditions given by
Eqs.~(\ref{eq:bc_d1}) and (\ref{eq:bc_d2}).

\section{Reflection coefficient\label{sec:refl} }

The plane wave solutions are assumed for the dynamic components of
the magnetization vector $m_{\alpha}(\mathbf{r},t)$: 
\begin{equation}
m_{\alpha}(\mathbf{r},t)=A_{\alpha}\text{e}^{i\left(\mathbf{k}_{\mathrm{i}}\cdot\mathbf{r}-\omega t\right)}+B_{\alpha}\mathrm{e}^{i\left(\mathbf{k}_{\mathrm{r}}\cdot\mathbf{r}-\omega t\right)},
\end{equation}
by substituting this solution to the boundary condition (\ref{eq:BCond})
we obtain: 
\begin{eqnarray}
 &  & \left[ik_{\mathrm{i},y}A_{\alpha}\mathrm{e}^{i\left(\mathbf{k}_{\mathrm{i}}\cdot\mathbf{r}-\omega t\right)}-ik_{\mathrm{r},y}B_{\alpha}\mathrm{e}^{i\left(\mathbf{-k}_{\mathrm{r}}\cdot\mathbf{r}-\omega t\right)}\right.\nonumber \\
 & + & \left.d\left(A_{\alpha}\mathrm{e}^{i\left(\mathbf{k}_{\mathrm{i}}\cdot\mathbf{r}-\omega t\right)}+B_{\alpha}\mathrm{e}^{i\left(\mathbf{-k}_{\mathrm{r}}\cdot\mathbf{r}-\omega t\right)}\right)\right]_{y=0}=0.
\end{eqnarray}
At the reflection point $k_{\mathrm{i},x}=k_{\mathrm{r},x}\equiv k_{x}$
and $k_{\mathrm{i},y}=-k_{\mathrm{r},y}\equiv k_{y}$ we get: 
\begin{equation}
ik_{y}A_{\alpha}-ik_{y}B_{\alpha}+d(A_{\alpha}+B_{\alpha})=0.
\end{equation}
Finally, the formula (\ref{eq:ReflCoefR}) for refractive index can
be easily derived: 
\begin{equation}
R\equiv\frac{B_{\alpha}}{A_{\alpha}}=\frac{ik_{y}+d}{ik_{y}-d}=\frac{i\sqrt{k^{2}-k_{x}^{2}}+d}{i\sqrt{k^{2}-k_{x}^{2}}-d}.
\end{equation}

\begin{acknowledgments}
We received funding from the European Union's Horizon 2020 research
and innovation programme under the Marie Sk{\l }odowska-Curie grant
agreement No 644348. P.G and M.K. acknowledge the financial assistance
from National Science Center of Poland (MagnoWa DEC-2-12/07/E/ST3/00538)
. Yu.S.D. and N.N.D. acknowledge Ministry of Education and Science
of Russian Federation (project No. 14.Z50.31.0015). K.Y.G. acknowledges
support by IKERBASQUE (the Basque Foundation for Science). The part
of calculations presented in this paper were performed at Poznan Supercomputing
and Networking Center. The authors would like to thank O. Yu. Gorobets
and Yu. I. Gorobets for fruitful discussion. \end{acknowledgments}


\begin{thebibliography}{99}
\bibitem{footnote_1} It is important to note an influence of used
in MMS finite difference (FD) cells dimensions on the obtained results,
especially when a surface magnetic anisotropy is included. Due to
specification of FD method used in MuMax3, where only volume anisotropy
$K_{\mathrm{V}}$ is included, surface anisotropy have to be introduced
in indirect way. We've assumed volume anisotropy in one row of finite
difference cells along edge of the film (at $y=0$). In this case
$K_{\mathrm{V}}(y)=\delta(0)K_{\mathrm{s}}/l_{y}$, where $\delta(y)$
is Dirac delta function. MMS confirmed that the size of cell $l_{x}\times l_{y}$
influences the results of simulations in the scale below the exchange
length. However, we have verified numerically, that in the performed
here MMS there is only slight change of the obtained values with further
decrease of the unit cell, i.e., for $2.5$~nm the convergence is
achieved.
\end{thebibliography}
\end{document}